# Flow field evolution and entrainment in a free surface plunging jet

Syed Harris Hassan[1], Tianqi Guo[2], Pavlos P. Vlachos[2,*]

[1] School of Engineering, RMIT University, 124 La Trobe St,
Melbourne VIC 3000, Australia

[2] School of Mechanical Engineering, Purdue University, 585 Purdue Mall,
West Lafayette, IN 47907, USA

**ABSTRACT**

We investigate ambient fluid entrainment and near-field flow characteristics of a free surface plunging jet for five Reynolds numbers ranging from 3000 to 10000 using time-resolved stereo particle image velocimetry (SPIV). We present time-averaged velocities, RMS velocity fluctuations, mean entrainment and unsteady flow features and compare them with previous studies on free jets. We find that plunging jets have a smaller potential core length, and earlier decay of the mean centerline velocity. The peak RMS velocity fluctuations occur at a location significantly upstream compared to the free jets reported in the literature. Near-field ambient fluid entrainment of plunging jets is measured for the first time and is found to be considerably higher than free jets in the low Reynolds number range. For the plunging jet case at Re = 3000, faster jet decay, higher levels of turbulent intensity in the near-field, and augmented mass entrainment result from strong primary vortices that give the turbulent/non-turbulent interface (TNTI) its convoluted shape which facilitates both bulk entrapment of ambient fluid and small scale nibbling because of larger surface area. These primary vortices occur right below the free surface and disintegrate into secondary structures at axial locations that are upstream compared to those of free jets. At higher Reynolds numbers, primary vortices are smaller in size, weak in swirling strength, and disintegrate prematurely, resulting in suppressed mixing and reduced entrainment efficiency.

[*] Author to whom correspondence should be addressed.
E-mail address: pvlachos@purdue.edu

## I. INTRODUCTION

A plunging jet consists of a free jet that exits from a nozzle and plunges into a quiescent pool of fluid from a certain height. This phenomenon is abundant in nature and is also encountered in various industrial processes. Hydraulic jumps and waterfalls are examples of this phenomenon occurring naturally whereby they contribute to river oxygenation, which is crucial for the underwater ecosystem. Industrial processes such as wastewater discharge, sump discharge, aeration of chemical reactions for gas absorption or mixing and pouring of molted metals and plastics involve plunging jets [1]. Such easy-to-generate flows are valuable owing to their mixing efficiency by exchanging mass, momentum, and heat. Studying these flows is therefore important in order to understand and control natural phenomena and design engineering applications. Moreover, plunging jets are important from the scientific standpoint as they involve interaction of a free jet with a gas-liquid interface, which generates a complex multi-phase, three-dimensional, and turbulent flow field that gives rise to numerous interesting questions [2].

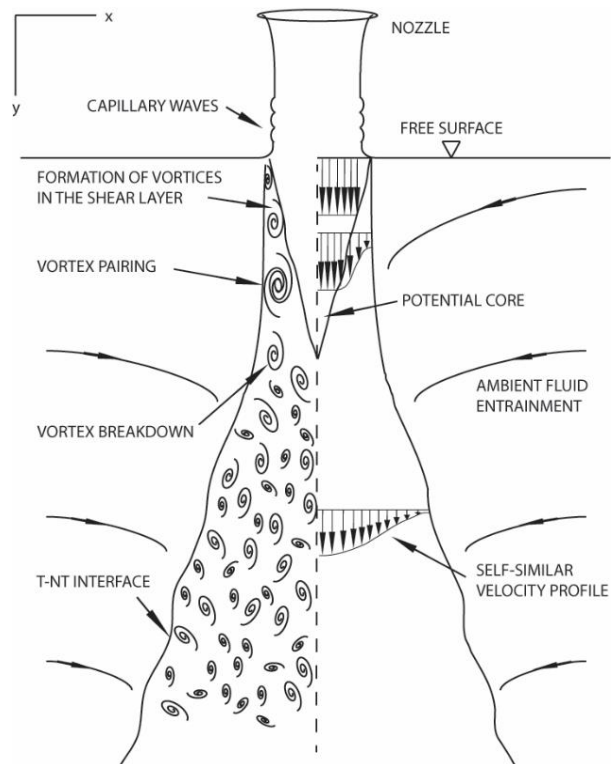

FIG. 1. Schematic diagram of a plunging jet.

FIG. 1 illustrates a plunging jet as it exits the nozzle. As the jet travels downwards, the liquid column undergoes acceleration due to gravity, resulting in an increase in the jet velocity and a decrease in its diameter [3]. Capillary waves form at the column base, followed by an increase in its diameter before impacting the free surface [4]. Beneath the free surface, the plunging jet evolves similar to a free jet whereby the axial velocity decreases with depth and the velocity profiles spread out due to diffused momentum and entrainment of ambient liquid. As the jet penetrates the free surface, velocity difference between the potential core and ambient fluid results in the roll up of the shear layer into primary vortical structures that resemble toroidal rings [5]–[9]. Formation of these vortices is due to growth of the Kelvin-Helmholtz instability [7]–[9] . As the primary vortices travel downwards, they grow in size and entrain ambient fluid towards the jet core resulting in its erosion and eventual termination at approximately $y/D = 4 - 5$ [7]–[11]. This growth is both linear and symmetrical both towards and away from the potential core. Apart from interacting with the potential core, primary vortices also interact with each other due to mutual induction and undergo pairing [5], [7], [9]. These pairing events happen single or multiple times depending on the Reynolds number [12], [13]. Studies have also shown secondary streamwise vortices or braids are responsible for further entrainment of ambient fluid towards the jet [8], [9], [14], [15]. With the end of the potential core, the centerline velocity decays resulting in weakened shear. Consequently, primary vortices become unstable and disintegrate into smaller secondary structures that move downstream leading to fully developed turbulent flow [6], [8].

Ambient fluid entrainment plays a crucial role in the decay and spreading of both free and plunging jets. The term entrainment comprises of all mechanisms through which irrotational ambient fluid is incorporated into the turbulent region [16]. One of the earliest studies to measure entrainment in free jets was by Ricou & Spalding in 1960 [17]. They designed a novel apparatus consisting of a porous walled cylindrical chamber surrounding the jet. Since the chamber length was comparable to the jet length, they established a mean value for the entrainment coefficient (the derivative of normalized entrainment rate) and reported an average value of $C_E = 0.32$. All their experiments were performed for $Re > 25000$, beyond which entrainment remained constant. Hill [18] modified the porous-walled apparatus by reducing the chamber height which enabled measurement of local entrainment rates by moving the chamber axially downstream from the nozzle. Experiments were performed for $Re > 60000$, and results indicated that the local entrainment coefficient started with a value of $C_E = 0.1$ at $y/D = 2$ from the nozzle and increased to $0.32$ at $y/D = 13$, which was in agreement with Ricou & Spalding [17]. Liepmann & Gharib [8] were the first to investigate this problem using Particle Image Velocimetry (PIV) and discussed the role of streamwise vortex pairs (braids) in the near-field entrainment of round jets. They showed that these braids develop from wave-like instabilities in the shear layer and play a major role in an inward flow of ambient fluid around the jet via streamwise vorticity. Han & Mungal [19] quantified ambient fluid entrainment in a free jet subjected to co-flow by direct PIV measurements. Although their entrainment coefficient agreed with Ricou & Spalding [17], it took 35 diameters to converge rather than 13 as reported by Hill [18]. El Hassan & Meslem [15] studied the effect of nozzle geometry on the near-field entrainment of a free jet at Re = 9500 through stereo PIV. Lobed and chevron nozzles were found to produce streamwise vortex more efficiently than round nozzles, resulting in much higher entrainment rates. This was in agreement with Liepmann & Gharib [8] that streamwise structures enhance near-field entrainment for free jets.

While our understanding of the flow characteristics and ambient fluid entrainment for free jets is comprehensive, similar questions for plunging jets have not been addressed in detail. Most of the literature regarding plunging jets has focused on air entrainment mechanisms [20]–[22], bubble size distribution [23], [24], bubble penetration depth [25], [26], mean residence time of air entrained [27], [28], and mixing characteristics [20], [22]. Very few studies have been dedicated to the flow field development of the plunging jet itself [29], [30], especially in the absence of air entrainment. And to the best of our knowledge, there has been no study on ambient fluid entrainment for plunging jets. Considering these research gaps, this work investigates the flow field of a plunging jet using time-resolved stereo particle image velocimetry (SPIV) at nozzle Reynolds number between 3000 and 10000. We quantify statistical flow properties such as mean centerline velocity decay, RMS velocity fluctuations and mean entrainment of plunging jets, and compare them to previous studies on free jets over a wide range of Reynolds numbers. In addition, the unsteady flow behavior involving vortex dynamics is also investigated and discussed.

## II. EXPERIMENTAL APPARATUS AND PROCEDURES

### A. Flow loop

Experiments are performed in a 0.3 m × 0.2 m × 0.3 m rectangular tank fabricated with clear acrylic sheets. A round and straight nozzle of inner diameter 7.8 mm is supported above the tank by an aluminum frame and carefully aligned perpendicular to the free surface by a spirit level. The height of the nozzle is adjusted at 2 nozzle diameters above the free surface. A gear pump (Model: C.P.78004-02) with pump head (Model: GB.P35.JSV.A) is used to generate a re-circulating flow. Experiments are done at five nozzle Reynolds numbers ($Re_N$ = 3000, 4000, 6000, 8000, and 10000). Within this range the pump gives a stable fluid column, and there are no entrained air bubbles appear below the free surface. FIG. 2 shows the schematic diagram of the setup.

### B. Time-resolved stereoscopic measurements

Neutrally buoyant spherical polystyrene particles (Duke Scientific, density of 1.05 g/cm$^3$) of 7 μm in diameter are homogeneously dispersed in the tank for flow seeding. Illumination is provided by a 527 nm, dual pulse Nd:YLF (neodymium: yttrium lithium fluoride) laser (Continuum Terra PIV 527-80-M). The laser beam is converted into a 1.5 mm thick light sheet using two spherical and one cylindrical lens. The particle-scattered light is recorded by two high-speed CMOS (complementary metal-oxide-semiconductor) cameras (Phantom Miro M340, Vision Research, 1300 × 2560 pixels) mounted at an angle of 30 degrees and at approximately 35 cm from the light sheet. Nikon objectives of 105 mm focal length and an aperture value of f# 8 are used with Schiempflug adapters [31]. The resulting field-of-view below the free surface is 40 mm by 80 mm. Both cameras and the laser are controlled by a synchronizer (TSI Laser Pulse Synchronizer 610036). For all tests, cameras are operated at 980 Hz while laser pulse rate is varied accordingly to achieve particle displacement of around 7 – 8 pixels per frame near the free surface.

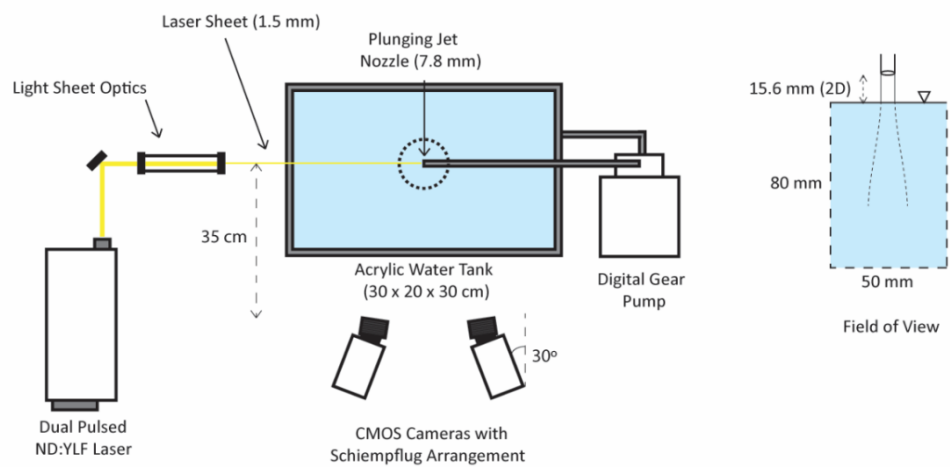

FIG. 2. Schematic diagram of the experimental setup.

Stereoscopic camera calibration is performed using a dual plane/dual side calibration plate with 1.5 mm dots spaced at 1 cm (TSI). The calibration target is translated in the water tank by a micrometer traverse such that 7 calibration planes are acquired by each camera (5 planes inside the laser sheet and 1 on either side) with 0.381 mm out-of- plane increments. A polynomial mapping function (cubic in-plane and quadratic out-of-plane) is then employed with a self-calibration [32] refinement using 200 images to improve measurement accuracy.

1200 image pairs are acquired for each test case, and then cross-correlated by LaVision DaVis 7.6 utilizing a multi-pass, iterative window deformation scheme with 32 × 32 window size and 50 % overlap for the first three passes and a subsequent 16 × 16 window size and 50 % overlap for the next three passes. After each pass, results were validated using velocity thresholding and Universal Outlier Detection [33] to replace bad vectors. The final grid size is 4 × 4 pixels giving 665 × 358 vectors in the field-of-view.

### C. Calculation of mean entrainment rate

Ambient fluid entrainment in jets can be directly measured through integration of mean velocity profiles obtained through PIV measurements [19], [34].

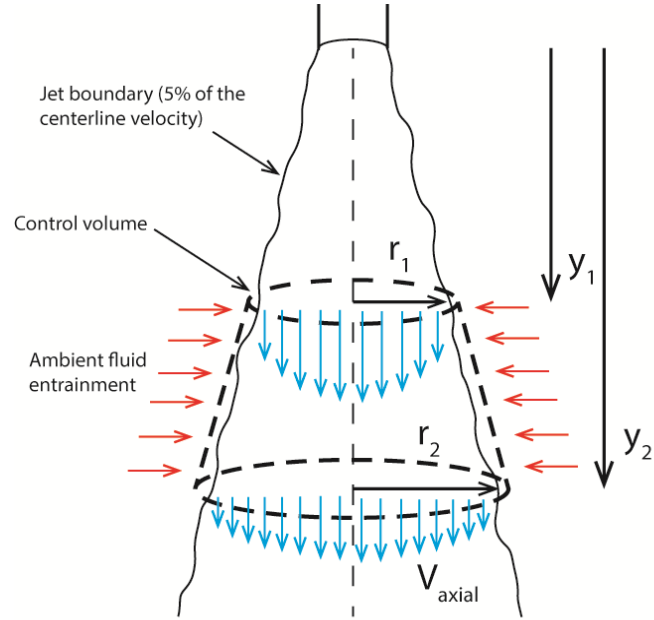

FIG. 3. Schematic diagram of the control volume used for entrainment evaluation from the axial velocity field.

FIG. 3 shows a schematic diagram of the control volume defined by axial locations $y_1$ and $y_2$ and radial locations $r_1$ and $r_2$. The red arrows show the ambient fluid being entrained by the jet while the blue arrows show the axial jet velocity. By mass conservation, the jet mass increase between $y_1$ and $y_2$ must be equal to the mass of the ambient fluid entrained radially through the control volume boundary. Mass flux entrained can then be calculated using the following equation

$$\dot{m}_{ent} = \int_0^{r_2} \rho \overline{V}(y_2, r) 2\pi r \, dr - \int_0^{r_1} \rho \overline{V}(y_1, r) 2\pi r \, dr, \qquad (1)$$

where ρ is the fluid density and $\overline{V}$ is the mean axial velocity obtained from 1200 snapshots of the velocity field. To use Equation (1), it is essential to define a criterion to identify the jet boundary in terms of the axial velocity. In this work, this criterion is defined as the radial position where the axial velocity of the jet has reduced to 5% of the centerline velocity. Values from Equation (1) can further be used to calculate the local jet entrainment coefficient as defined by Ricou and Spalding [17]

$$C_E = \frac{d_0}{\dot{m}_0}\frac{d\dot{m}}{dx}, \qquad (2)$$

where $\frac{d\dot{m}}{dx}$ represents the entrainment rate (the entrained mass flux per unit axial distance), and it is normalized by the jet diameter ($d_0$), and the initial mass flow rate ($m_0$). In this study, we calculate both the normalized mass flux and entrainment rate for all test cases.

### D. Scaling of the plunging across the air-water interface.

In studies on free jets, radial and axial distances are normalized by the nozzle diameter ($D_N$) which is constant for all Reynolds numbers. However, in plunging jets, the liquid column diameter downstream of the nozzle is a function of both the axial distance and Reynolds number. FIG. 4 displays relationship between the jet diameters (normalized by the nozzle diameter) before and after impact as the Reynolds number is increased from 3000 to 10000.

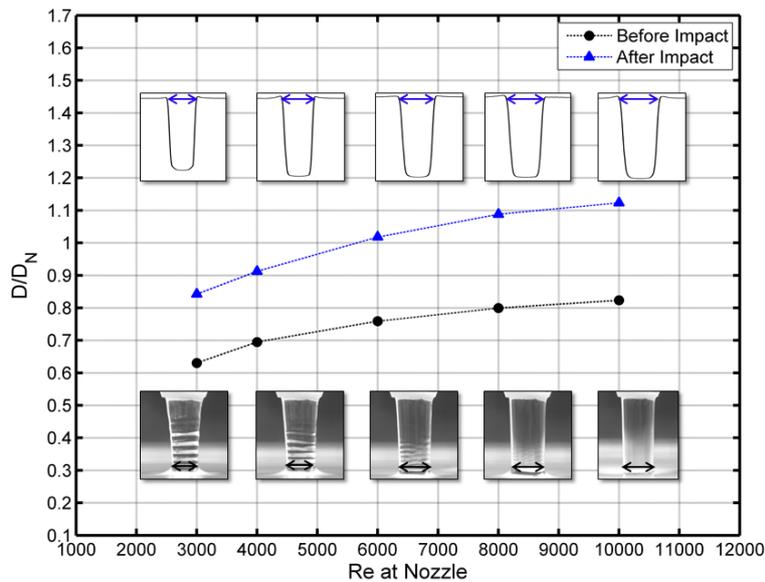

FIG. 4. Relationship between the plunging jet diameter and Reynolds number before and after impact.

Lower insets show the measurement location of the jet column diameter at the narrowest point above the meniscus before impact. Magnification calibration for this measurement is performed by placing a measurement scale in air besides the jet column. Post-impact plunging jet diameter is acquired by measuring the width of the axial velocity profiles right below the free surface (upper insets), and magnification value is acquired from Stereo PIV calibration. As the Reynolds number is increased, post-impact jet diameter increases from 0.85 at $Re_N = 3000$ to 1.12 at $Re_N = 10000$. Moreover, increase in the plunging jet diameter across the free surface is approximately 33–36% for all Reynolds numbers, which necessitates revision of length and velocity scaling. This revision is essential in order to compare our results with previous studies on free jets as it allows representing plunging jets with different nozzle velocities, nozzle diameters and plunging heights as free jets at the air-water interface with a certain diameter ($D_{FS}$) and velocity ($V_{FS}$). We introduce a new free surface Reynolds number ($Re_{FS}$) in Equation (3)

$$Re_{FS} = \frac{\rho D_{FS} V_{FS}}{\mu}, \qquad (3)$$

with values for each case shown in Table I.

Table I. Scaling of the plunging jet using parameters at the free surface

| Nozzle Reynolds number $Re_N$ | Diameter below the free surface $D_{FS}$ (m) | Velocity below the free surface $V_{FS}$ (m/s) | Free surface Reynolds number $Re_{FS}$ |
|---|---|---|---|
| 3000 | 0.0065 | 0.70 | 5004 |
| 4000 | 0.0071 | 0.78 | 6086 |
| 6000 | 0.0079 | 0.98 | 8774 |
| 8000 | 0.0084 | 1.17 | 10628 |
| 10000 | 0.0087 | 1.37 | 13292 |

The physical quantities presented in this paper are therefore reported in non-dimensional units divided by the jet diameter $D_{FS}$ and jet velocity $V_{FS}$ at the free surface.

## III. RESULTS

### A. Centerline velocity decay

Decay of mean centerline velocities of the plunging jet is illustrated in FIG. 5 (color symbols). Axial distance below the free surface and streamwise velocities are normalized by jet diameter ($D_{FS}$) and maximum streamwise velocity ($V_{FS}$) at the free surface. Representative results from the literature regarding decay of free jets are also shown (open symbols).

The axial distance where the mean streamwise velocity starts to decay is usually referred to as the length of the potential core, which is a conical region between the nozzle (or free surface in case of a plunging jet) and the closure point of shear layers [6]–[8], [10]. For free jets in the literature, centerline velocity starts to decay in the range y/D = 5 – 7 [7], [10], [13], [35]–[38], which is in sharp contrast to the plunging jets herein, where the centerline velocity in the potential core remains above $V/V_{FS}$ = 0.98 until approximately $y/D_{FS}$ = 2 – 2.5 for all cases except for $Re_{FS}$ = 5004. For the latter case, an initial descent at $y/D_{FS}$ = 1 is followed by a plateau where $V/V_{FS}$ stays at 0.96 before the final decay starts at $y/D_{FS}$ = 2.5. This behavior contradicts literature results at higher Reynolds numbers, where the centerline velocity stays constant until the decay region starts [7], [13], [35], [39]. However, a recent study by Todde et al. [10] on free jets of Re = 1620 – 4050 reports a similar plateau in the initial region with periodic velocity modulation by mutual induction and pairing of strong ring vortices in this region. In later sections, we discuss similar primary vortices at $Re_{FS}$= 5004, a phenomenon significantly weaker for higher Re cases.

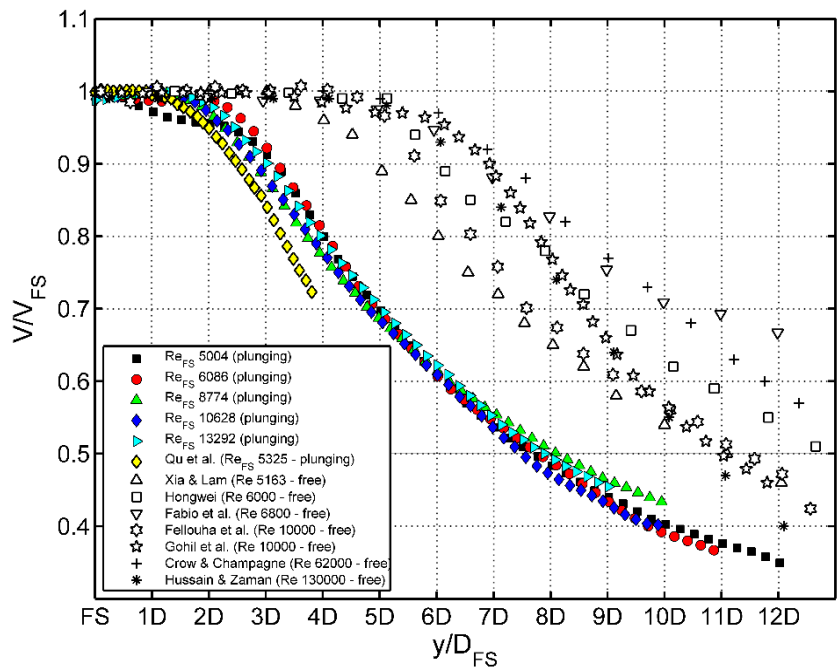

FIG. 5. Mean centerline velocity decay of the plunging jet cases normalized by the jet velocity at the free surface ($V_{FS}$). Representative results in the literature regarding free jets.

Beyond the potential core, velocity profiles of plunging jets decay significantly faster than free jets for all Reynolds numbers, which is indicated by the position where $V/V_{FS}$ = 0.5. For plunging jets, this location is approximately 7.5 – 8 $D_{FS}$ below free surface, while for free jets this location lies between 11 – 13 D. A similar premature decay of the centerline velocity was also reported by Qu et al. [30] (yellow symbols in FIG. 5) at Re = 5325 and a plunging height of y/D = 2 (same as our study). Early cessation of the potential core and rapid decay of the centerline velocity indicates faster development of plunging jets and therefore makes them less efficient than free jets in conserving momentum.

### B. RMS of centerline velocity fluctuations

FIG. 6 shows the RMS of jet centerline velocity fluctuations normalized by $V_{FS}$ for the axial, radial and azimuthal directions respectively (color symbols) while representative results from literature regarding axial RMS fluctuations of free jets are shown in FIG. 6 (a) (black symbols). Peak centerline RMS velocity fluctuation have been associated with rise in turbulent activity due to closure of the shear layers and start of turbulent mixing throughout the jet [6], [7], [13], [39]. Similar to the trend of mean centerline velocity decay, peak RMS values in plunging jets appear significantly upstream ($y/D_{FS} = 4 – 5$) of the values reported for free jets in the literature ($y/D \sim 9$). This again demonstrates faster decay of plunging jets compared to free jets. The discrepancy in the initial values of RMS velocity fluctuations near $y/D_{FS} = 0$ between free and plunging jets may be due to the plunging effect. For the plunging jet cases, significantly higher levels of axial and radial velocity fluctuations are observed for the $Re_{FS} = 5004$ case. Axial velocity fluctuations ($v/V_{FS}$) increase rapidly from 0.04 near the free surface and reach a peak of 0.16 at $y/D_{FS} = 4$, thereafter decreasing gradually to 0.08 at $y/D_{FS} = 12$ (FIG. 6(a)). Similar pattern holds for $Re_{FS} = 6086, 8774, 10628$ and $13292$ cases, although their peak values are lower ($v/V_{FS} = 0.14$) and occur near $y/D_{FS} = 4.5$. Further downstream, the amplitudes decay in a linear fashion suggesting fully developed turbulence.

Centerline radial velocity fluctuations ($u/V_{FS}$, FIG. 6(b)) show a similar trend as axial fluctuations, with slightly lower magnitudes. At $Re_{FS} = 5004$, $u/V_{FS}$ rises rapidly from 0.02 near the free surface to 0.15 at $y/D_{FS} = 4$, followed by swift decrease to 0.10 at $y/D_{FS} = 7$ and finally a gradual decay to 0.06 at $y/D_{FS} = 12$. At $Re_{FS} = 6086$, $u/V_{FS}$ attains a peak of 0.12 while in the latter three cases ($Re_{FS} = 8774 – 13292$), peak values are slightly lower at 0.11, followed by a linear decay to 0.075 at $y/D_{FS} = 10$. Azimuthal velocity fluctuations ($w/V_{FS}$, FIG. 6(c)) have the same range as both axial and radial cases, however the deviation of $Re_{FS} = 5004$ case is not as pronounced with peak activity occurring at 0.135.

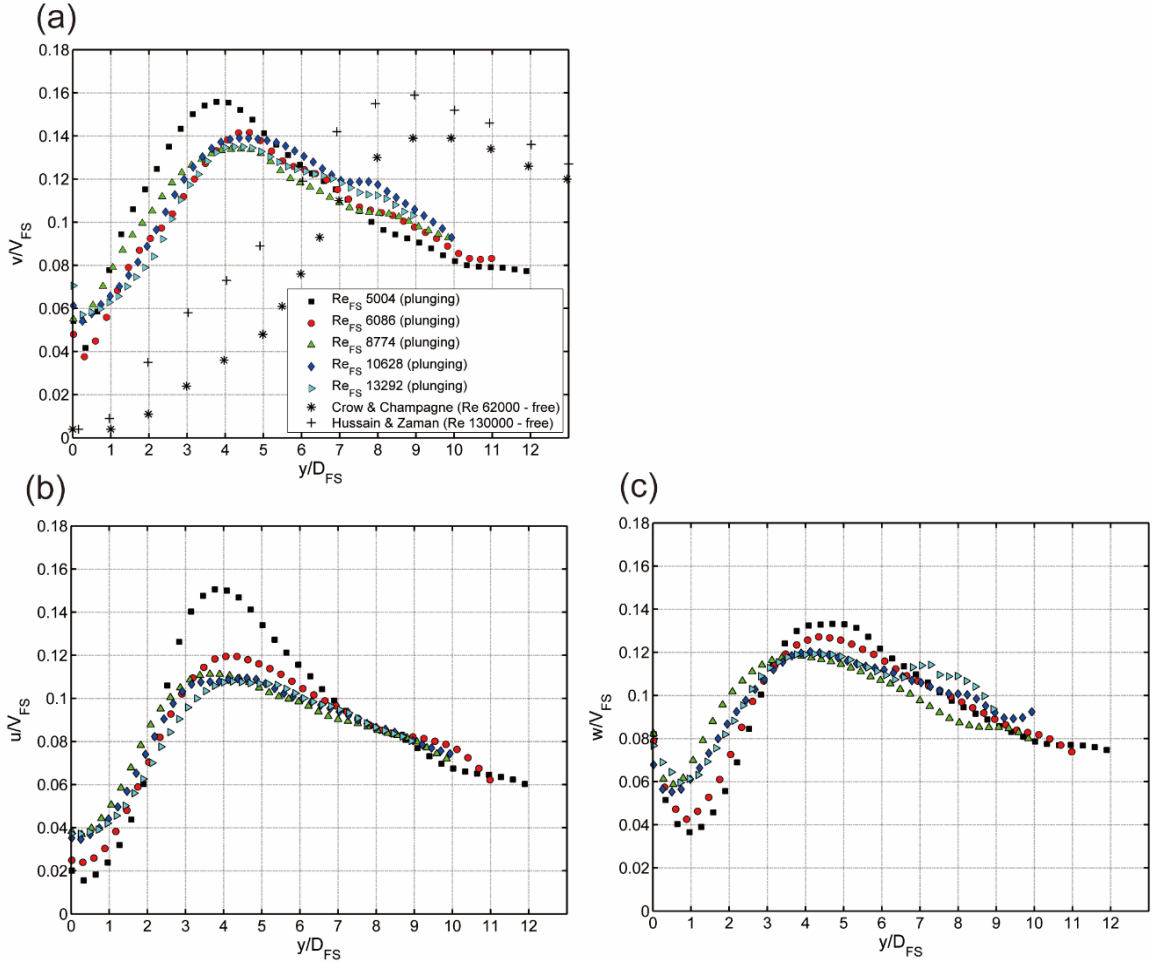

FIG. 6. RMS of centerline velocity fluctuations for (a) axial, (b) radial and (c) azimuthal directions compared with literature results on free jets.

The deviation of the $Re_{FS}$ = 5004 case in both axial (by 15 %) and radial (by 35 %) RMS velocity fluctuations is in sharp contrast to previous investigations of free jets at high Reynolds numbers (FIG. 8 in Crow & Champagne [7]; Re: 62000 – 124000 and FIG. 2 in Hussain & Zaman [13]; Re: 32000 – 113000), where an order of magnitude change in the Reynolds number have little effect on the peak values. The behavior at $Re_{FS}$ = 8774, 10682 and 13292 appears to be more consistent with the previous studies as velocity fluctuations become independent of the Reynolds number and collapse on top of each other. In a recent study however, Reynolds number dependence of RMS velocity fluctuation also exists for Re ~ 1000 – 6000 [10]. As will be discussed later, location of high RMS velocity fluctuations at $Re_{FS}$ = 5004 is consistent with the region where strong primary vortices disintegrate into small secondary structures, causing increase in the turbulent intensity and mixing.

## C. Entrainment of ambient fluid

FIG. 7(a) illustrates the mass entrainment ratio ($m/m_0$) obtained by averaging 1200 snapshots of axial velocity profiles integrated according to Equation (1). Entrainment for all plunging jet cases stays close to the results of Ricou & Spalding [17] for the first 5 diameters, after which values start to diverge with the $Re_{FS}$ = 5004 case rising steeply to $m/m_0$ = 5 at $y/D_{FS}$ = 9, indicating that it has entrained 5 times its initial mass at a depth of 9 diameters. At the same depth, $Re_{FS}$ = 6080 has entrained approximately 4.5 times its initial mass. Mass ratio profiles for the $Re_{FS}$ = 8774, 10628 and 13232 cases collapse onto each other with $m/m_0$ = 4 at $y/D_{FS}$ = 9.

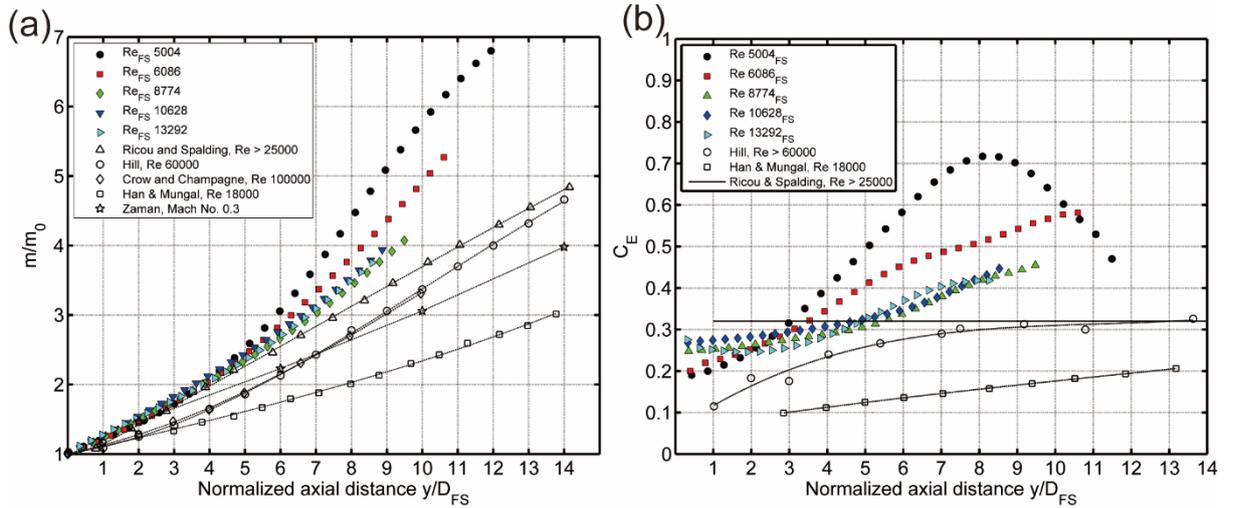

FIG. 7. a) Mean entrainment rate ($m/m_0$) and b) entrainment coefficient ($C_E$) of plunging jets compared with literature results on free jets.

In FIG. 7(b) the entrainment coefficient defined by Equation (2) is plotted for comparison with literature results on free jets. Entrainment coefficient ($C_E$) quantifies the local rate at which mass is being entrained when compared to its initial mass flow rate ($m_0$). Entrainment coefficient obtained by Ricou & Spalding [17], Hill [18] and Han & Mungal [19] involving free jets all achieve a constant value of 0.32 in the far field, indicating a steady state has been reached. For plunging jet at $Re_{FS}$ = 5004, $C_E$ starts with 0.2 near the free surface and undergoes a sharp increase at $y/D_{FS}$ = 3, reaching 0.72 at $y/D_{FS}$ = 8, followed by a descent to approximately 0.5 at $y/D_{FS}$ = 12. The $C_E$ values for $Re_{FS}$ = 6068 show a gradual growth from 0.2 near the free surface to 0.45 at $y/D_{FS}$ = 6, thereafter growing linearly to approximately 0.6 at $y/D_{FS}$ = 11. Similar to the mass entrainment ratio profiles, $C_E$ values of $Re_{FS}$ = 8774, 10628 and 13292 cases collapse onto each other, starting with a higher value of approximately 0.27 near the free surface and reaching 0.45 at $y/D_{FS}$ = 9. Collapse of ($m/m_0$) and $C_E$ profiles beyond $Re_{FS} \sim 9000$ demonstrates that plunging jet entrainment has become independent of the Reynolds number. This is consistent with results of Ricou & Spalding [17] who reported that the ratio $m/m_0$ became constant for free jets beyond $Re = 25000$. However, for plunging jets, Reynolds number independence occurs much earlier at $Re_{FS} \sim 9000$. Due to limited field of view in the axial direction, $C_E$ values for all cases do not approach an asymptotic value, however, it has been established in numerous studies that the entrainment coefficients in all jets, irrespective of their initial conditions, do approach an asymptotic value when the flow achieves fully developed turbulence [8], [16], [18], [19], [40]. Overall, plunging jets at low Reynolds numbers are much more efficient in entraining ambient fluid in the near-field, and increase in the Reynolds number causes a decrease in the mean entrainment coefficient.

### D. Unsteady flow behavior

The instantaneous flow organization of plunging jets is shown in FIG. 8 for $Re_{FS} = 5004$, 8774 and 13292 cases in each row. The temporal sequence (separated by 3Δt) consists of velocity vectors normalized by the jet axial velocity at the free surface ($V_{FS}$), with color contours representing the normalized swirling strength ($\lambda_{ci}$) calculated according to Zhou et al. [41]. The close-up views of the shear layer from the windows in FIG. 8 (b, e and h) are shown in FIG. 9.

FIG. 8(a – c) shows three snapshots at $Re_{FS} = 5004$. Strong primary vortices convect along the shear layer, growing in size as they entrain ambient fluid before becoming unstable and disintegrating into smaller secondary structures. Primary vortices induce a strong inwards flow at their trailing edges along which ambient fluid is forced towards the jet core (red arrows in FIG. 9 (a)). This inwards flow is called an entrainment zone or wedge and is inclined at an angle of approximately 45º from the jet axis [6]. Between $y/D_{FS} = 0.5 – 1.2$, two adjacent vortices undergo pairing (arrows A, FIG. 8 (a – c)), which occurs earlier than free jets in previous studies over a similar Reynolds number range. Leipman & Gharib [8] reported that the first pairing event for a water jet at $Re = 5500$ usually takes place before 2.5 jet diameters while a recent study by Violato & Scarano shows occurrence between 2.7 – 3.5 diameters [9]. Pairing of the adjacent vortices in the shear layer occurs as a result of mutual induction whereby the downstream vortex induces an inwards and axial flow at its trailing edge, causing an increase in the convective velocity of the upstream vortex. On the other hand, an opposite but less pronounced flow is induced by the upstream vortex at its leading edge as it pushes the flow outwards, causing a decrease in the convective velocity of the upstream vortex [9]. Eventually, as a consequence of the difference in the convective velocity of the two adjacent vortices, a pairing event occurs. Subsequently, the resulting single vortex increases in size and convects downstream until $y/D_{FS} = 2.5 – 3.0$ before disintegrating into smaller secondary structures (arrows B, FIG. 8(a – c)).

For the latter cases at $Re_{FS} = 8774$ and $13292$, primary vortices are smaller in size and have low swirling strength compared to the previous case. Overall arrangement of the shear layer is disorganized with numerous smaller vortical structures accompanying the primary vortices as they convect downstream. Often a "vortex train" appears below the free surface where primary vortices interact and then quickly disintegrate into smaller secondary structures. Smaller primary vortices with lower swirling strength result in weak and disorganized entrainment zones (red arrows in FIG. 9(b) and (c)). In comparison with the previous case at $Re_{FS} = 5004$, both pairing and disintegration events have shifted upstream to $y/D_{FS} = 0.3 – 0.5$ and $y/D_{FS} = 0.5 – 1$, respectively (arrows C and D in FIG. 8 (d – h))

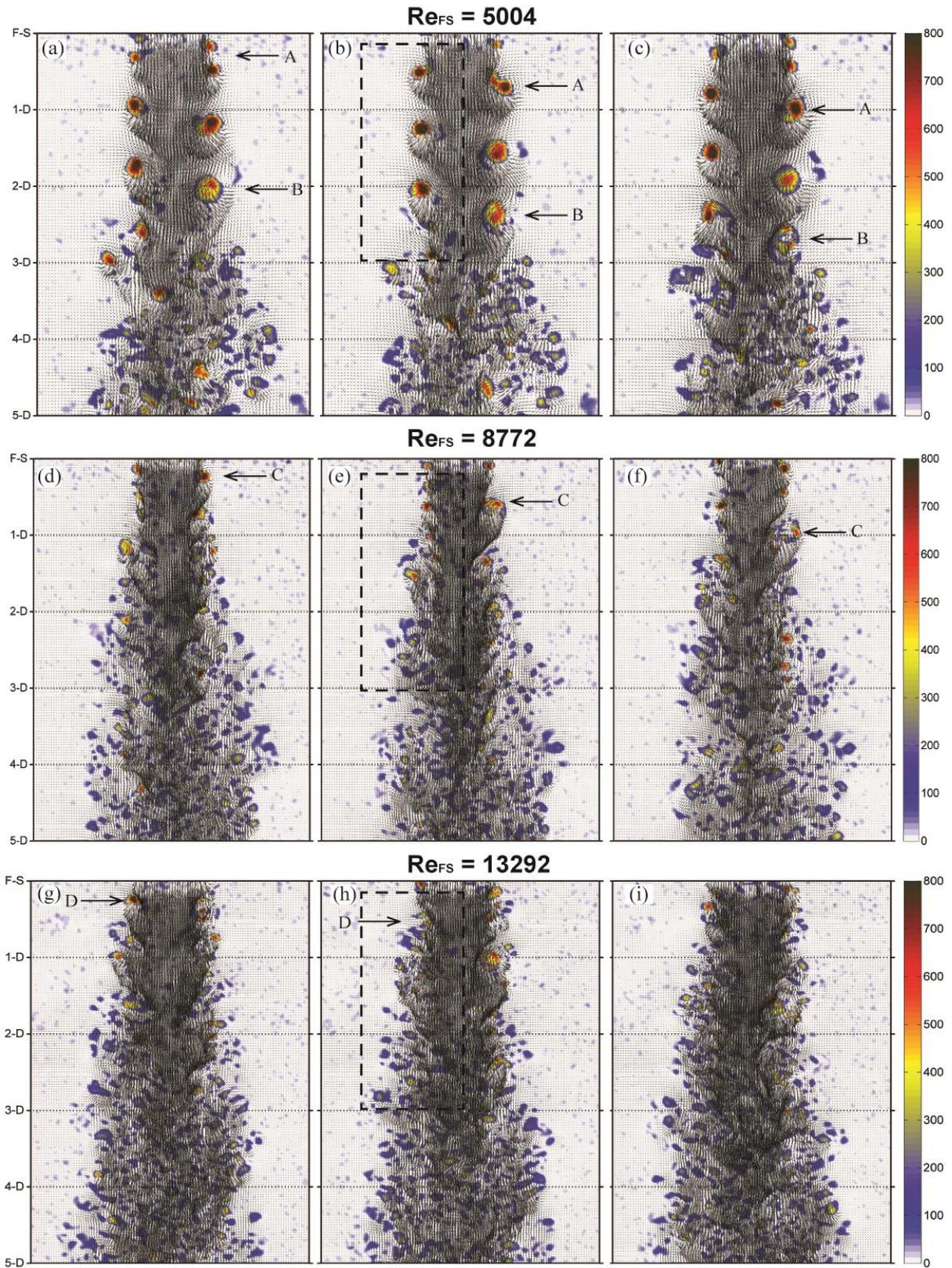

FIG. 8. Time sequence visualization of the plunging jet at $Re_{FS}$ = 5004, 8772 and 13292. Iso-contours represent swirling strength of coherent structures ($\lambda_{ci}$). Time separation between snapshots: $3\Delta t$. The dashed windows (black) show spatial locations plotted in FIG. 9

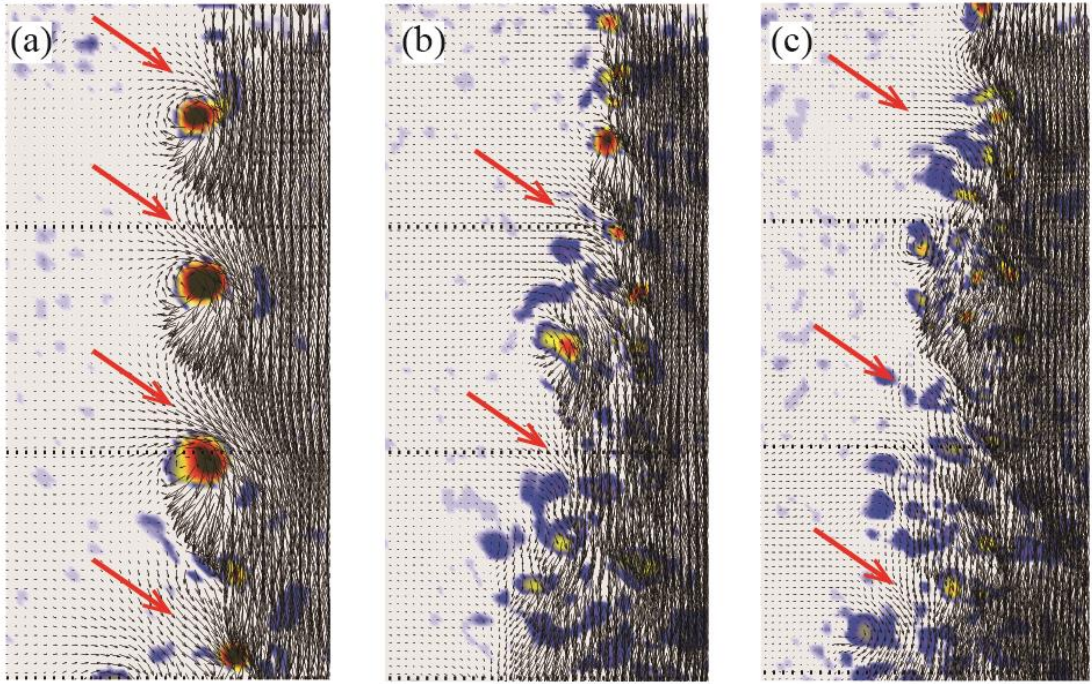

FIG. 9. Close-up view of the initial shear layer of plunging jets at (a) $Re_{FS} = 5004$ (b) $Re_{FS} = 8774$ and (c) $Re_{FS} = 13292$ taken from FIG. 8 (black dashed windows). Red arrows show entrainment zones besides primary vortices.

## IV. DISCUSSION

In this study, statistical flow properties of plunging jets such as mean centerline velocity decay, RMS velocity fluctuations and unsteady flow behavior involving vortex dynamics are analyzed from SPIV measurements. Moreover, we have provided the first comparison study on mean entrainment of ambient fluid between plunging and free jets.

### A. Comparison between plunging jet and free jet

For all plunging jet cases in this study, primary vortices appear immediately below the free surface. This is in contrast with previous studies regarding free jets where it takes approximately 2 – 2.5 nozzle diameters for the vortices to shed from the shear layer in the same Reynolds number range [8]–[10], [38]. Although the mechanism for shear layer roll–up is the same, i.e. velocity difference between the high speed jet column and the ambient quiescent fluid [2], [8], the plunging effect accelerates this roll–up, resulting in early vortex development that facilitates bulk entrainment of the ambient fluid by vortex induction, leading to early initiation of the entrainment zone. This in turn causes premature erosion of the potential core and subsequent decay of the centerline velocity. For plunging jets between $Re_{FS} = 5004 - 13292$, centerline velocity starts decaying at $y/D_{FS} = 2 - 2.5$, while previous studies show that for free jets, this decay usually starts at $y/D_{FS} = 5 - 7$ over a wide range of

Reynolds numbers between 5000 and 130000 (FIG. 5). Similar trend is seen for RMS velocity fluctuations along the centerline where peak values for plunging jets occur at $y/D_{FS}$ = 4 – 5 while previous studies on free jets show that peak fluctuations occur at $y/D$ = 8 – 9 [7], [13] (FIG. 6(a)). This was also concluded by Qu et al [30] in their study comparing the flow structure of a plunging and submerged jet at Re = 5325 where increased RMS fluctuations and an early start of the entrainment zone led to premature decay of the mean centerline velocity (Fig. 14 in the cited paper).

Early decay of plunging jets is also linked to higher near-field entrainment for all Reynolds numbers. At $Re_{FS}$ = 5004, the entrainment coefficient $C_E$ of 0.7 at $y/D_{FS}$ = 8 is more than twice the value reported by Ricou & Spalding for free jets at Re > 25000. At the same axial distance, plunging jet at $Re_{FS}$ = 6086 has $C_E$ = 0.5, while entrainment coefficients for the latter three cases collapse along the axial length and reach a value of 0.4. This demonstrates that with increasing Reynolds number, mean entrainment reduces and eventually becomes invariant at $Re_{FS}$ ~ 9000. Similar behavior was also reported by Ricou & Spalding for free jets, albeit at Re > 25000 (Fig. 3 of the cited paper). Due to limited field of view, entrainment coefficients for all cases did not achieve an asymptotic value, which is reported to be 0.32 for free jets.

### B. Increased entrainment efficiency for low Re plunging jets

Among the plunging jets, the $Re_{FS}$ = 5004 case has higher levels of RMS velocity fluctuations which leads to higher turbulent intensity and enhanced mixing that facilitates mass and momentum transfer from the shear layers towards the jet centerline, resulting in higher entrainment efficiency. This relationship was initially explored by Crow & Champagne [7] in their study on excitation of free jets. They demonstrated that forcing the jet at various excitation frequencies inflated the turbulent intensities along the centerline in the near-field, resulting in increased levels of entrainment (Fig. 29 of the cited paper). Higher levels of entrainment have also been correlated with increased turbulent intensity in a recent study by Quinn et al. [42] on circular and triangular jets in which they reported that latter were able to entrain a greater amount of fluid owing to higher turbulent intensity along the centerline (Fig. 6 and 9 of the cited paper).

Higher levels of RMS velocity fluctuations that lead to increased entrainment efficiency are due to the presence of strong primary vortices in the shear layer at $Re_{FS}$ = 5004, a phenomenon that is significantly weak in the latter cases. As these vortices convect downstream, they entrap large amount of ambient fluid and eventually disintegrate at y/$D_{FS}$ = 2.5 – 3 into smaller secondary structures. The location where disintegration occurs (arrows B in FIG. 8) agrees well with the initiation of centerline velocity decay (FIG. 5) for $Re_{FS}$ = 5004. This agreement was also reported by Leipmen & Gharib [8] for a free jet at Re = 5500. They termed this as mixing transition and explained that disintegration occurs because the velocity difference between the ambient fluid and high speed core decreases, attenuating the shear that supports the vortices. For the latter cases at $Re_{FS}$ =8774 and 13292, the location where vortices disintegrate at (y/$D_{FS}$ = 0.5 – 1) does not coincide with the location of the centerline velocity decay (y/$D_{FS}$ ~ 2). The reason for this discrepancy lies in the mechanisms of vortex decay and destruction. At low Reynolds numbers, vortices generated in the shear layer can grow until the end of potential core [8], [10], [11]. On the contrary, at high Reynolds numbers, the development of the shear layer is rapid which results in quick decay of the vortical structures and subsequent generation of fine-scale turbulence that suppresses large-scale mixing. This is in agreement with observations of Kim & Choi [11] in their LES study and Todde et al. [10] in their experimental study on the structure of low Reynolds number free jets.

Recently, several high resolution computational and experimental studies on free jets and boundary layers have performed a rigorous analysis on the characteristics of mass entrainment at the turbulent/non-turbulent interface (TNTI) [43]–[46]. By means of mass-flux spectra, it has been shown that entrainment is a two-stage process that occurs on two different scales with substantial scale separation. The first stage involves "bulk entrapment" of fluid across the TNTI by large eddies of the same scale as the jet diameter. These eddies also give the TNTI its characteristic shape that consists of indentations on a variety of macro-scales. The second stage involves "nibbling" that occurs on the order of Taylor microscale and is independent of the large-scale motions [43]. Nibbling is carried out by a thin viscous "superlayer" that resides on the large scale indentations and has high levels of vorticity [43]–[46]. In this two-stage process, the large eddies are responsible for a) bulk entrainment of ambient non-turbulent fluid across the interface and b) providing a larger surface area to the TNTI so that the second stage nibbling process occurs more efficiently [44]. Applying this analogy to the plunging jet cases, we believe that higher entrainment efficiency at $Re_{FS}$ = 5004 occurs because of large primary and secondary vortical structures that are responsible for the convoluted shape of the TNTI facilitating both bulk entrapment of the ambient fluid and small scale nibbling due to larger surface area. At higher Reynolds numbers, primary and secondary vortices saturating the TNTI are smaller with less swirling strength, resulting in small scale indentations and an overall reduced surface area. This suppresses bulk entrapment of ambient fluid and the small-scale nibbling process that is dependent on the TNTI area, leading to overall reduced entrainment efficiency.

V. CONCLUSION

The flow field evolution and ambient fluid entrainment in a plunging jet is investigated between nozzle Reynolds numbers of 3000 and 10000 (free surface Reynolds numbers of 5004 and 13292) with time-resolved stereo particle image velocimetry (SPIV). The plunging effect alters the flow field evolution by changing the unsteady flow structure beneath the free surface. Primary vortical structures are formed right below the free surface resulting in an early initiation of entrainment. Early formation of primary vortices further leads to early vortex pairing and breakdown, resulting in premature closure of shear layers and cessation of the potential core. This is also evident from the statistical flow properties where the locations of centerline velocity decay and peak activity in RMS velocity fluctuations occur approximately 4 – 5 jet diameters upstream of the same locations for free jets in previous studies.

This study presents the first effort to quantify ambient fluid entrainment in plunging jets. Due to early development, plunging jets exhibit higher levels of entrainment in the near-field compared to free jets. Entrainment values decrease with an increase in Reynolds number and become independent of the same beyond $Re_{FS} = 9000$. Early development and higher levels of entrainment also show that plunging jets are not as efficient as free jets in conserving momentum and therefore have better mixing characteristics.

Future work may include extending the measurement domain in the axial direction and using Tomographic PIV to fully resolve the three-dimensional near-field flow structure. This would provide more accurate fluid entrainment results by directly integrating the whole three-dimensional flow field rather than assuming axisymmetry.